\documentclass[conference]{IEEEtran}
\usepackage{cite}
\usepackage{amsmath,amssymb}
\usepackage{graphicx}
\usepackage{booktabs,multirow}
\usepackage{textcomp}
\usepackage{xurl}
\usepackage[hidelinks]{hyperref}
\hypersetup{pdftitle={VerfGraph: Efficient Zero-Knowledge Verifiable Graph Query Evaluation System}}
\begin{document}
\title{VerfGraph: Efficient Zero-Knowledge Verifiable Graph Query Evaluation System}
\author{
\IEEEauthorblockN{
Tianyi Ma\IEEEauthorrefmark{1},
Yuncong Zhang\IEEEauthorrefmark{2},
Yuanzhuo Yu\IEEEauthorrefmark{1},
Boshi Yuan\IEEEauthorrefmark{1},
and Shi-Feng Sun\IEEEauthorrefmark{1}}
\IEEEauthorblockA{
\IEEEauthorrefmark{1}Shanghai Jiao Tong University\qquad
\IEEEauthorrefmark{2}Shandong University\\
\{mty021129, yzyu2000, nemoyuan2008, shifeng.sun\}@sjtu.edu.cn,
yuncong@sdu.edu.cn}
}
\maketitle

\begin{abstract}
Graph databases are widely used to manage relationship-rich data and support multi-hop traversal and shortest-path queries in applications such as financial risk control and social networks. When a client issues such queries to a database owner, the returned result may be incorrect. However, the client cannot directly verify it because the graph is typically large and private, making re-executing the query both infeasible and insecure. Verifiable and confidential graph query evaluation thus plays an important role in such applications.

We present VerfGraph, a zero-knowledge verifiable graph query evaluation system, which enables a client to verify that the query result was correctly computed over a committed graph, without revealing anything beyond the public inputs and what is implied by the query result. While adopting the expansion-centric operator model introduced by ZKGraph (ICDE’26), VerfGraph improves both proving efficiency and query expressiveness in two key ways. First, we propose new proof methods for the core operators in ZKGraph, such as graph expansion and shortest path, that require substantially fewer expensive comparisons and eliminate edge canonicalization for bidirectional edge relations. Second, we introduce dedicated proofs for count and weighted shortest-path operators, which ZKGraph does not support, broadening the class of graph queries that can be verified. We evaluate VerfGraph using the LDBC Social Network Benchmark. Compared with the state-of-the-art zero-knowledge graph query system ZKGraph on complex IC queries, VerfGraph achieves 33.9×–59.7× speedups in proof generation and reduces peak prover memory by a factor of 4.5×–6.6×, while maintaining millisecond-level verification and producing sub-MB proofs.
\end{abstract}


\section{Introduction}
Graph databases organize data around entities and relationships, making them well suited to workloads dominated by neighborhood exploration, multi-hop connectivity, and path analysis~\cite{angles2008survey,angles2017foundations,erling2015ldbc}. In outsourced or cross-organizational deployments, however, the party that evaluates a query may not be fully trusted. A service provider may evaluate a query against an unintended or stale snapshot, omit qualifying vertices or paths, or return an incorrect result because of implementation faults or adversarial behavior. Because the client neither sees the private graph nor can necessarily afford to re-execute an expensive traversal, the query result alone does not establish that it was computed faithfully.

This setting gives rise to two coupled requirements. The first is \emph{confidentiality}: the raw graph, sensitive node and edge properties, and private intermediate results should remain hidden except for information revealed by an authorized query answer. The second is \emph{verifiability}: the client should be able to establish that the reported answer is correct and complete for a precisely specified query and a public commitment to a graph snapshot. Neither requirement implies the other. Revealing the graph enables direct recomputation but defeats confidentiality, whereas keeping it private without cryptographic evidence forces the client to trust an opaque execution.

For example, consider an e-commerce platform that maintains a private graph whose vertices represent products, sellers, and buyers, and whose edges encode transactions, co-purchase relationships, and referral links. An external auditor may be authorized to ask for the number of buyers reachable from a suspicious seller within a fixed number of hops and the minimum-risk referral path between two flagged accounts, where edge weights encode risk scores. Disclosing the full graph would reveal sensitive commercial and user data, yet returning only the count and path gives the auditor no basis for detecting omitted edges, use of an outdated snapshot, or an incorrectly computed path. A suitable protocol should bind each answer to a committed graph snapshot and prove the requested properties without exposing unrelated accounts, transactions, or intermediate traversal states.

Zero-knowledge proofs (ZKPs) provide a cryptographic foundation for
addressing this problem. A ZKP allows a prover to convince a verifier that a
public statement is true without revealing the private witness beyond what is
implied by the statement~\cite{goldwasser1985knowledge}.
To use ZKPs to establish query correctness, existing ZK database
systems commonly decompose queries into operators and design
specialized proof constructions for their input--output relations.
Proving these input--output relations efficiently is a central
challenge: the proofs must establish that each operator's output
is correct and complete without revealing private intermediate values.
For example, ZKSQL~\cite{li2023zksql} and
PoneglyphDB~\cite{gu2024poneglyphdb} develop such proof constructions
for relational operators, including sorting, joins, and aggregation.
However, graph queries introduce proof requirements, such as establishing
shortest-path optimality and correctly handling traversal over bidirectional
edge relations, for which these systems do not directly provide specialized
proof constructions.
To address these requirements, ZKGraph~\cite{wu2025zkgraph} introduces an
expansion-centric model for decomposing queries into composable
operators, and develops specialized proof constructions for its core
graph operators, including single-source expansion, set-based expansion,
and single-source shortest path (SSSP).
Through these operator-proof constructions, ZKGraph demonstrates
the effectiveness of using zero-knowledge proofs to support
confidential and verifiable graph query evaluation.

Despite this progress, two issues remain to be addressed.
First, ZKGraph's constructions for set-based expansion and
bidirectional edge relations rely on repeated
comparisons.
Proving these comparisons requires a large number of range checks
to establish whether one value is greater than another, and
performing these checks across graph records increases
proof-generation costs.
However, simply removing these range checks would no longer
guarantee operator correctness, so new proof constructions
are needed to preserve correctness while reducing the number
of range checks.
Second, ZKGraph does not provide dedicated proof constructions
for count or weighted SSSP, which are important graph
query operations, and developing such constructions would broaden
the query support of zero-knowledge verifiable graph query systems.

To address these issues, we present VerfGraph, an efficient
zero-knowledge verifiable graph query evaluation system.
VerfGraph retains ZKGraph's expansion-centric model and advances
the operator design in two ways.
First, for set-based expansion, we replace ZKGraph's construction,
which locates the source of each edge within the sorted source set,
with a partition-based proof. For expansion and SSSP over
bidirectional edge relations, instead of canonicalizing each edge
by reordering its endpoints, we treat the edge relation as two
directed views and verify both traversal directions directly.
Together, these changes reduce the required sorting and comparison
checks, thereby lowering range-check overhead while preserving
correctness and completeness.
Second, we introduce a dedicated proof construction for the
count operator and extend the SSSP construction to weighted
graphs, broadening the class of graph queries that can be verified.

We implement VerfGraph's operator proofs as AIR
circuits~\cite{bensasson2018stark} using Plonky3~\cite{plonky3} with a
polynomial commitment scheme (PCS) based on Fast Reed--Solomon
Interactive Oracle Proofs of Proximity
(FRI)~\cite{bensasson2018fri}.
Unlike ZKGraph's Halo2/KZG backend~\cite{halo2,kate2010commitments},
this transparent, hash-based backend requires no trusted setup
and relies on assumptions believed to resist quantum
attacks~\cite{bensasson2018stark}.
We evaluate VerfGraph using
the LDBC SNB benchmark~\cite{ldbcSNBDocs}.
Compared with ZKGraph on complex IC queries, VerfGraph
substantially reduces proof-generation time and peak prover
memory while retaining millisecond-level verification
and sub-MB proofs.

The remainder of this paper is organized as follows. Section~\ref{sec:background} introduces the necessary background. Section~\ref{sec:system-overview} presents the system overview. Section~\ref{sec:circuit-design} describes the circuit design of the verifiable graph operators. Section~\ref{sec:experiments} presents our experimental results. Section~\ref{sec:related-work} discusses related work, and Section~\ref{sec:conclusion} concludes the paper.

\section{Background}
\label{sec:background}

\subsection{Zero-Knowledge Proofs}

Zero-knowledge proofs (ZKPs) allow a prover to convince a verifier that
a statement is true without revealing private information beyond what
the statement reveals~\cite{goldwasser1985knowledge}.
In graph query processing, the prover is the server holding the private
graph, and the verifier is the client requesting a query.
For example, the server may prove that a returned set contains exactly
the neighbors of a queried vertex without disclosing the rest of the
graph. Here, the statement concerns the correctness of the
returned neighbors, while the private graph records and intermediate
values used to establish it form the witness.

For practical query verification, a particular class of ZKPs, called
zero-knowledge succinct non-interactive arguments of knowledge
(zkSNARKs), provides three properties: proofs are compact; verification
is fast relative to the computation being proved; and proofs can be
checked without further interaction with the prover. These properties
make zkSNARKs suitable for verifiable graph queries: the server returns
the result with a compact proof, and the client checks it
efficiently without accessing the private graph, repeating the query
computation, or further communicating with the server.

\subsection{AIR Arithmetization}

To prove query correctness with a zkSNARK, we first express the
conditions for a valid result as polynomial equations, called
constraints. This translation, known as arithmetization, allows the
verifier to check a proof that these equations hold without repeating
the query computation.

VerfGraph uses an Algebraic Intermediate Representation
(AIR)~\cite{bensasson2018stark}, which organizes the values in
these equations into a table called an execution trace. Each row records
a computation step, while columns store inputs and intermediate
values. These polynomial constraints enforce the computation within each
row, connect successive rows, and require relevant trace values to
match the public inputs and claimed output. Satisfying these constraints
establishes that the trace represents a valid computation producing the
claimed result.

\begin{figure}[!t]
  \centering
  \includegraphics[width=0.94\linewidth]{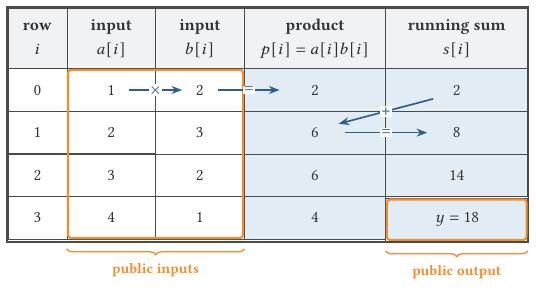}
  \caption{Four-row multiplication-and-accumulation AIR trace.}
  \label{fig:air-multiply-accumulate}
\end{figure}

\textbf{Example:} To illustrate how these constraints establish
correctness, consider computing the sum of pairwise products of two
vectors, as shown in Figure~\ref{fig:air-multiply-accumulate}.
Let the public inputs be $a=(1,2,3,4)$ and $b=(2,3,2,1)$, and let the
claimed output be $y=18$. Each row stores two inputs $a[i]$ and $b[i]$,
their product $p[i]$, and a running sum $s[i]$.
A local polynomial constraint requires each product to be computed
correctly: $p[i]-a[i]b[i]=0$ for $0\leq i\leq3$.
To accumulate these products, a transition constraint connects
consecutive rows:
$s[i+1]-s[i]-p[i+1]=0$ for $0\leq i<3$.
The boundary constraints $s[0]=p[0]$ and $s[3]=y$ initialize the sum
and require its final value to equal the claimed output.
With $a$ and $b$ fixed to the public inputs, these constraints require
$p=(2,6,6,4)$ and $s=(2,8,14,18)$.

The prover uses the trace as a witness to generate a zkSNARK proving
that these constraints hold without revealing private trace values.
The same row-oriented representation applies to database scans,
accumulations, and iterative graph operations, where repeated steps
can be described by the same constraints across rows.

\subsection{Permutation, Lookup, and Range Checks}
\label{sec:prelim-checks}

This subsection introduces three protocols that are repeatedly used in our
graph query proofs: permutation, lookup, and range checks. At the database
level, permutation checks establish multiset (bag) equality, lookup checks
establish table membership, and range checks support integer comparisons.
Each protocol is implemented by a proof circuit, which encodes the conditions
that the prover's data must satisfy. We analyze the circuit complexity of
each protocol in terms of circuit size, measured by the total number of
constraints enforced across the execution trace.
Proof-generation cost depends on circuit size and generally increases with it.

\emph{Permutation check.} A permutation check proves that two length-$N$
columns $A$ and $B$ contain the same values with the same multiplicities,
regardless of order. The column values belong to a finite field $\mathbb{F}$,
a finite set supporting addition, subtraction, multiplication, and division
by nonzero values. To compare the columns without matching rows individually,
we follow LogUp and the reciprocal argument of
Bulletproofs++~\cite{habock2022multivariate,eagen2024bulletproofspp} and compare
sums of reciprocals. With a random challenge $\beta\in\mathbb{F}$ sampled
after the columns are determined, the protocol checks whether
\[
  \sum_{i=0}^{N-1}\frac{1}{\beta+A[i]}
  =
  \sum_{i=0}^{N-1}\frac{1}{\beta+B[i]}.
\]
Except with negligible probability over the random choice of $\beta$, this
equality holds if and only if $A$ and $B$ represent the same multiset.
The circuit complexity of this protocol is $O(N)$.

\emph{Lookup check.} A lookup check proves that every value in a column
$A\in\mathbb{F}^N$ appears in a reference table $B\in\mathbb{F}^\ell$.
A reference entry may appear multiple times in $A$. The prover records
the multiplicity of each $B[j]$ in $A$ as $m[j]$ and uses these
counts as weights in the reciprocal sum above. With a random challenge
$\beta\in\mathbb{F}$, the protocol checks whether
\[
  \sum_{i=0}^{N-1}\frac{1}{\beta+A[i]}
  =
  \sum_{j=0}^{\ell-1}\frac{m[j]}{\beta+B[j]}.
\]
The circuit complexity of this protocol is $O(N+\ell)$.

Both checks above operate on single columns, whereas graph relations such as
edge tables span multiple columns. To extend permutation and lookup checks
to multi-column rows, we compress each row tuple $(t_1,\ldots,t_k)$ into a
single field element via a random linear combination:
\[
  t_1+\alpha_1 t_2+\cdots+\alpha_{k-1}t_k,
\]
where $\alpha_1,\ldots,\alpha_{k-1}\in\mathbb{F}$ are random challenges.
Except with negligible probability, distinct tuples do not collide under this
compression, allowing permutation and lookup checks on the compressed column
to verify multi-column relations.

\emph{Range check.} Finite-field arithmetic does not directly express integer
ordering, so we verify integer comparisons by bounding differences between
the compared values. For integers $x$ and $y$, checking $x\le y$ amounts to
checking that the difference $y-x$ is nonnegative. A range check establishes
such bounds by proving that a value represents an integer in $[0,2^w)$,
where $w$ is the required bit width. A lookup against $\{0,\ldots,2^w-1\}$
would establish this bound but require a table with $2^w$ entries. The
protocol instead splits the value into smaller bit groups, called limbs,
and checks that they reconstruct the original value. Each limb is then
checked against a smaller lookup table. Using
8-bit limbs requires the table $\{0,\ldots,255\}$; using 16-bit limbs reduces
the number of limbs but requires a table with 65,536 entries. Following
ZKGraph's implementation, VerfGraph uses 8-bit limbs,
requiring $n=\lceil w/8\rceil$ limb lookup checks to range-check a column
of $w$-bit values. If $w$ is not divisible
by 8, the most significant limb is checked against a smaller table covering
only the remaining bits.
For a length-$N$ column, the circuit complexity of this protocol is
$O(n(N+T))$, where $T=256$ is the lookup-table size.
Because each range check splits a value into $n$ limbs and performs a
lookup check per limb, it is substantially more expensive than a single
permutation or lookup check over columns of the same length, and becomes a
major performance bottleneck for comparison-heavy operators.

\subsection{Graph Database Operators and Edge Storage Formats}
\label{sec:graph-query-primitives}

Graph queries compose fundamental operators such as neighborhood traversal,
property filtering, shortest-path finding, and counting.
Because these operators involve iterative, data-dependent traversals over
irregular structures, directly encoding their execution into static proof
circuits is prohibitively expensive. To address this challenge,
ZKGraph~\cite{wu2025zkgraph} introduces an expansion-centric model that
decomposes queries into composable operators (such as single-source
expansion, set-based expansion, and SSSP) and verifies
each operator's input--output relation while enforcing consistency across
intermediate relations. VerfGraph adopts this model, designs new proof
constructions for core operators such as expansion and shortest path, and
extends it to new operations.

The cost of proving these operators also depends on how graph edges are
represented. Two common representations are compressed sparse row (CSR),
which stores each vertex's neighbors consecutively in an array and uses
offsets to locate each neighbor list, and the edge-list format, which stores
each edge as a row containing its two endpoints and optional attributes (such
as edge weights). ZKGraph shows that the edge-list format avoids the costly offset-indexing
constraints of CSR and yields substantially faster proof generation. VerfGraph
therefore adopts the edge-list representation. Under this representation, an edge relation is
either directed, where each row $(u,v)$ can be traversed only from $u$ to
$v$, or bidirectional, where each edge is stored once as $(u,v)$ but can be
traversed in either direction.

\section{System Overview}
\label{sec:system-overview}

\subsection{System Model}
\label{sec:system-model}

VerfGraph consists of two parties: a prover $\mathcal{P}$ and a verifier
$\mathcal{V}$. The prover maintains a private graph database $G$, including
its vertex and edge relations and their attributes. Given a graph query $q$
submitted by $\mathcal{V}$, $\mathcal{P}$ evaluates $q$ over $G$ and returns
the query result $y$ together with a proof $\pi$ of the computation.

Upon receiving $y$ and $\pi$, $\mathcal{V}$ checks $\pi$ using the public query
description, the claimed result, and the corresponding verification parameters.
If verification succeeds, $\mathcal{V}$ accepts $y$ without accessing the
private graph or re-executing the query.

\subsection{Security Model}
\label{sec:security-model}

VerfGraph considers two potential adversaries. A malicious prover may
substitute graph data or construct incorrect intermediate and final query
results. A malicious verifier may attempt to learn private graph records or
intermediate values beyond what is revealed by the public inputs and query
result. To fix the graph data before a query is issued, the prover computes
and publishes an immutable database commitment $C_G$ offline, and the verifier
checks the received proof directly against $C_G$. Under this model, we
consider four properties:

\begingroup
\setlength{\leftmargini}{1.2em}
\begin{itemize}
\item \textbf{Completeness.} An honest prover that correctly evaluates the
declared query over the graph data committed by $C_G$ can generate a proof
accepted by the verifier.
\item \textbf{Soundness.} The probability that a malicious prover convinces
the verifier to accept a result inconsistent with the query over the
graph data committed by $C_G$ is negligible.
\item \textbf{Knowledge soundness.} When the verifier accepts a proof, the
prover possesses a valid witness for evaluating the declared query over the
graph data committed by $C_G$.
\item \textbf{Zero knowledge.} The verifier learns no information about
private graph records, intermediate relations, or auxiliary witness values
beyond what is implied by the public inputs and query result.
\end{itemize}
\endgroup

We next explain how the proof construction and operator constraints support
these properties.

\subsubsection*{Database and computation integrity}
To bind query evaluation to the private graph $G$, the prover encodes
database columns as polynomials and commits to them using a polynomial
commitment scheme (PCS), which binds a polynomial to a compact digest $C_G$
without revealing its values and later verifies its evaluations via opening
proofs. During proof generation, the prover commits to the execution-trace
polynomials and generates opening proofs, with random challenges derived
non-interactively via the Fiat--Shamir heuristic~\cite{fiat1986prove}. Checking
these openings against $C_G$ and the trace commitments allows the verifier to
confirm that the database values in the proof are consistent with the
committed columns. Starting from these values, AIR constraints establish each
operator's input--output relation, require its output to match the
next operator's input, and bind the final output to the public result $y$,
ensuring that $y$ follows from evaluating $q$ over the committed graph.

To instantiate the PCS, two major approaches are group-based schemes,
such as KZG~\cite{kate2010commitments} and IPA~\cite{bootle2016efficient}, and hash-based schemes. Hash-based schemes require no
trusted setup, rely on collision-resistant hash
functions~\cite{merkle1987digital,bensasson2016iop}, and are believed to resist
known quantum attacks~\cite{bensasson2018stark}. These properties motivate our
choice of an FRI-based PCS~\cite{bensasson2018fri}, a representative hash-based
construction.

\subsubsection*{Witness privacy}
VerfGraph protects private witness values via a randomized proof
construction following Hab{\"o}ck and Al Kindi~\cite{habock2024adding}, which
blinds the execution trace and committed polynomials with random values so
that proof verification reveals no private entries. Beyond individual entries,
graph operators produce intermediate relations whose sizes depend on
private data. To prevent these sizes from changing public trace dimensions,
VerfGraph follows ZKSQL~\cite{li2023zksql} and pads relations with dummy rows
to predetermined public capacities. Private Boolean activity tags distinguish
actual rows from dummy rows and prevent dummy rows from affecting query
results. With these padded relations, each operator applies the same constraint
pattern across a fixed-capacity trace, yielding an oblivious AIR layout for
fixed public inputs.

\subsubsection*{Public information and scope}
The query, query plan, result, database commitment, circuit description, and
trace dimensions and capacities are public. The graph schema, relation and
attribute types, and edge directionality are also public information needed to
interpret the query. Given this public information, our security model
addresses database and computation integrity against a malicious prover and
witness privacy against a malicious verifier. Other concerns, including
query-result inference, access control, denial of service, and side channels in
native query execution and witness construction, are outside our scope.

\subsection{System Workflow}
\label{sec:system-workflow}

\begin{figure}[!t]
  \centering
  \includegraphics[width=0.90\linewidth]{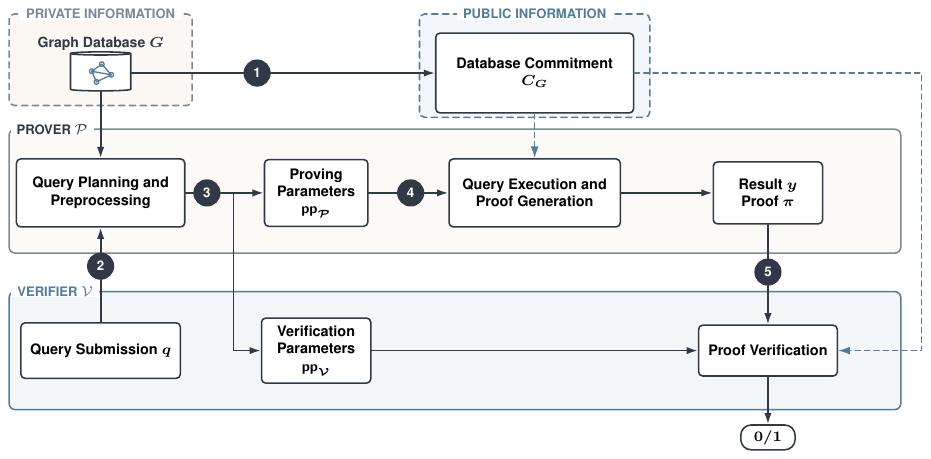}
  \caption{System workflow of VerfGraph.}
  \label{fig:system-workflow}
\end{figure}

VerfGraph processes a graph query in five stages, as summarized in
Figure~\ref{fig:system-workflow}.

\textbf{Offline database commitment.}
The prover encodes the graph-data columns as polynomials and commits
to these polynomials using the PCS, producing $C_G$. The verifier records
$C_G$ for subsequent verification, fixing the graph data against which the
query result must be proved.

\textbf{Query submission.}
With $C_G$ recorded, the verifier submits a query $q$ to be evaluated
over the committed graph data.

\textbf{Query planning and preprocessing.}
Following the expansion-centric model (Section~\ref{sec:graph-query-primitives}),
VerfGraph decomposes $q$ into graph operators and organizes them into a query
plan. From this plan, it constructs a proof circuit in AIR form by combining
the operators' constraints and connecting their input and output relations.
Preprocessing this circuit produces proving parameters
$\mathsf{pp}_{\mathcal{P}}$ for proof generation and verification parameters
$\mathsf{pp}_{\mathcal{V}}$, which the verifier uses together with the public
circuit description to check proofs.

\textbf{Query execution and proof generation.}
The prover executes the query plan over its private graph data to obtain the
result $y$. From the database values and intermediate operator relations,
it constructs a witness for the proof circuit. Using this witness and
$\mathsf{pp}_{\mathcal{P}}$, the prover generates a proof $\pi$ that $y$ is the
result of evaluating $q$ over the graph data committed by $C_G$. The prover
returns $y$ and $\pi$ to the verifier.

\textbf{Proof verification.}
The verifier checks the received proof $\pi$ using the public query
description, the recorded commitment $C_G$, the claimed result $y$, and
$\mathsf{pp}_{\mathcal{V}}$. If verification succeeds, the verifier accepts
$y$; otherwise, it rejects the response.

\section{Circuit Design}
\label{sec:circuit-design}

Following the expansion-centric model described in
Section~\ref{sec:graph-query-primitives}, this section presents VerfGraph's
circuit designs for core operators, including single-source expansion,
set-based expansion, single-source shortest path (SSSP), and count.
For these operators, we present circuit constructions using the protocols
introduced in Section~\ref{sec:prelim-checks} and analyze their circuit complexity.
We then compare the resulting circuit complexity with that of ZKGraph in
Section~\ref{sec:cost-comparison}. For clarity, we present each operator's
constraints over its active rows, while activity tags gate these constraints in
the padded traces (Section~\ref{sec:security-model}).

\subsection{Single-Source Expansion}
\label{sec:single-source-expansion}

Single-source expansion is a basic traversal operation in graph query
processing. Given a source vertex, it returns the one-hop edges that can
be traversed from that vertex, and therefore serves as a building block
for graph queries. The input consists of a source identifier $id_s$ and
an edge-list table $D_e$. The output table $D_{out}=(C,D)$ contains
exactly the rows $(id_s,id_t)$ for which $id_t$ is reachable from
$id_s$ in one hop through an edge recorded in $D_e$.

Which input edges belong in the output depends on whether $D_e$ is
directed or bidirectional. We consider the two cases
separately.

\subsubsection*{Circuit design for the directed edge relation}
For a directed edge relation, let $D_e=(A,B)$, where $A$ stores source
identifiers and $B$ stores target identifiers. We follow the same
proof-construction strategy as ZKGraph: selection followed by multiset
matching. Selection identifies exactly the input edges whose source is
$id_s$, while multiset matching requires $D_{out}$ to contain all and only
these edges with the same multiplicities.

To prove the selection, the prover introduces a flag
column $is_{id}$ over $D_e$, where
$is_{id}[i]=1$ iff $A[i]=id_s$, and $is_{id}[i]=0$ otherwise. The flag
is proved using an inverse-witness column $inv_{id}$, where
$inv_{id}[i]=(id_s-A[i])^{-1}$ if $A[i]\neq id_s$, and $inv_{id}[i]=0$
otherwise.
For every $i\in\{0,\ldots,|A|-1\}$, the circuit constrains
$is_{id}[i]$ to be Boolean and enforces
\[
  is_{id}[i](id_s-A[i])=0,
  \qquad
  is_{id}[i]+inv_{id}[i](id_s-A[i])=1.
\]
When $A[i]=id_s$, the second constraint forces $is_{id}[i]=1$; when
$A[i]\neq id_s$, the first constraint forces $is_{id}[i]=0$, and the
second verifies the inverse witness. The remaining task is to prove
that the selected edge rows are exactly the output rows. Using a random
challenge $\alpha$, we compress each row into a single field element as
$C_{input}[i] = A[i] + \alpha B[i]$ and
$C_{output}[i] = C[i] + \alpha D[i]$.
With another random challenge $\beta$, the prover uses a permutation check
to prove that the rows of $D_e$ satisfying $is_{id}[i]=1$ and the rows of
$D_{out}$ are permutations of one another, namely,
\[
  \sum_{i=0}^{|A|-1}
  \frac{is_{id}[i]}{\beta+C_{input}[i]}
  =
  \sum_{i=0}^{|C|-1}
  \frac{1}{\beta+C_{output}[i]}.
\]
These constraints establish the correctness of single-source expansion over
directed relations.

The circuit complexity of this operator comes from checking the source-selection
flags over the $E$ input edges, with circuit complexity $O(E)$, and matching the
selected edges with the $R$ output edges through a permutation check, with
circuit complexity $O(E+R)$. Since $R\leq E$, these constraints yield an
overall circuit complexity of $O(E)$.

\subsubsection*{ZKGraph's circuit design for the bidirectional edge relation}
For a bidirectional edge relation, let $D_e=(A,B)$, where each row stores two
endpoints and denotes a symmetric adjacency that can be traversed in
either direction. Applying the directed protocol only to the first column
would prove only the edges where $A[i]=id_s$ and would miss valid
traversals where $B[i]=id_s$, thereby failing to establish completeness.

\begingroup
\emergencystretch=1em
Duplicating each edge in both directions during preprocessing would require
committing $2E$ edge rows and slow down proof generation. To avoid this
overhead, ZKGraph introduces Bidirectional Relationship Canonicalization
(BiRC)~\cite{wu2025zkgraph}. Because a traversal $(id_s,id_t)$ in $D_{out}$
may be stored as either $(id_s,id_t)$ or $(id_t,id_s)$ in $D_e$, BiRC gives
each edge a unique representation by canonicalizing each row of $D_e$ and
$D_{out}$ so that the smaller endpoint appears first. It enforces sum and
product equality to check that each canonicalized row contains the same two
endpoints as the original row, and uses two range checks, one for each table,
to prove that the smaller endpoint appears first. With both tables in
canonical form, it marks the rows of $D_e$ that contain $id_s$ and uses a
permutation check to match the selected rows with the canonicalized
$D_{out}$.\footnote{The ZKGraph paper presents BiRC and states that
it can be integrated into expansion and SSSP over bidirectional relationships.
The procedures described in this section follow its public implementation at
\url{https://github.com/Hao8172/ZKGraph}.} These two range checks require
multiple limb lookups per row, adding to the circuit complexity.
\par\endgroup

\subsubsection*{Our circuit design}
For expansion over bidirectional relations, we avoid both this $2E$-row
commitment overhead and BiRC's canonicalization checks by treating $D_e$ as
two directed views, $(A,B)$ and $(B,A)$, without duplicating its rows. For single-source expansion,
we select rows $(A[i],B[i])$ with $A[i]=id_s$ from the first view and rows
$(B[i],A[i])$ with $B[i]=id_s$ from the second. In both cases, $id_s$ appears first.
The selected rows from both views therefore represent all traversals from
$id_s$, and a single permutation check proves that $D_{out}$ contains
exactly these rows with the same multiplicities.

To identify the selected rows in each view, the prover provides two Boolean
flag columns $left_{equal}$ and $right_{equal}$, where $left_{equal}[i]=1$
iff $A[i]=id_s$ and $right_{equal}[i]=1$ iff $B[i]=id_s$ (and $0$
otherwise). The prover proves both flag columns using the same
inverse-witness equality constraints as in the directed design. The
$left_{equal}$ flags identify forward traversals, and the $right_{equal}$
flags identify reverse traversals.
To match both directions against the output, we compress the rows with a
random challenge $\alpha$ as $C_{input}[i] = A[i] + \alpha B[i]$,
$C'_{input}[i] = B[i] + \alpha A[i]$, and
$C_{output}[i] = C[i] + \alpha D[i]$.
With another random challenge $\beta$, the circuit checks the
equality:\footnote{If self-loops are
permitted, the second numerator can be gated by $1-left_{equal}[i]$ to
avoid counting the same loop twice; our benchmark relations do not contain
such rows.}
\begingroup
\thinmuskip=2mu \medmuskip=2mu \thickmuskip=3mu
\[
  \sum_{i=0}^{|A|-1}\!
  \left(
    \frac{left_{equal}[i]}{\beta+C_{input}[i]}
    +
    \frac{right_{equal}[i]}{\beta+C'_{input}[i]}
  \right)
  =
  \sum_{i=0}^{|C|-1}\!
  \frac{1}{\beta+C_{output}[i]}.
\]
\endgroup

These constraints establish the correctness of single-source expansion over
bidirectional relations.

The circuit complexity of this operator comes from checking the two source-selection
flags for each of the $E$ input edges, with circuit complexity $O(E)$, and a single
permutation check over both orientations of the input edges and the $R$
output edges, with circuit complexity $O(E+R)$. Since $R\leq E$, these
constraints yield an overall circuit complexity of $O(E)$.

\subsubsection*{Correctness}
For both constructions, correctness follows from two facts. First, the flag constraints
identify exactly the edge rows that can be traversed from $id_s$. In the
directed case this means rows with $A[i]=id_s$; in the bidirectional
case it means rows with either endpoint equal to $id_s$. Second, the
permutation check proves that the selected traversals and the rows of
$D_{out}$ are permutations of one another. Together, they
prove the correctness of the one-hop traversals from $id_s$.

\subsection{Set-Based Expansion}
\label{sec:set-based-expansion}

The set-based expansion operator generalizes single-source expansion
from one source vertex to a set of source vertices. Its input consists
of a source set $ID_s$ and an edge-list
table $D_e$. The output table $D_{out}=(C,D)$ contains exactly the
one-hop traversals starting from a vertex in $ID_s$, where $C$ and $D$ store
the starting and adjacent vertex identifiers.

We consider set-based expansion over a directed edge relation in this
subsection. Unless stated otherwise, ``set-based expansion''
hereafter refers to this directed case. Let $D_e=(A,B)$, where $A$ and $B$
store source and target identifiers.

\subsubsection*{ZKGraph's circuit design}
To avoid repeating single-source expansion for each vertex in $ID_s$, ZKGraph
proves whether each edge source belongs to $ID_s$ by locating it between two
consecutive values in the sorted source set: the source belongs to $ID_s$
exactly when it equals the lower value.

To construct these intervals, ZKGraph augments $ID_s$ with minimum and maximum
boundary values outside the identifier domain and sorts the resulting column
as $\widetilde{ID}_s$. It also
sorts $D_e$ by source identifier to obtain
$\widetilde D_e=(\widetilde A,\widetilde B)$. The first permutation check
binds $\widetilde{ID}_s$ to the augmented source set, and the second binds
$\widetilde D_e$ to $D_e$. Two range checks then enforce
$\widetilde{ID}_s[i] < \widetilde{ID}_s[i+1]$ and
$\widetilde A[i] \leq \widetilde A[i+1]$ on adjacent rows, establishing
both sort orders.

For every sorted edge source $\widetilde A[i]$, the prover supplies a lower
bracket $L[i]$, equal to the greatest value in $\widetilde{ID}_s$ not
exceeding $\widetilde A[i]$, and its successor $U[i]$. A lookup check against
the consecutive-pair columns derived from $\widetilde{ID}_s$ proves that
$(L[i],U[i])$ are adjacent values in $\widetilde{ID}_s$. The remaining two range checks
enforce $L[i]\leq\widetilde A[i]$ and $\widetilde A[i]<U[i]$. Hence
$\widetilde A[i]$ belongs to $ID_s$ exactly when
$\widetilde A[i]=L[i]$. A Boolean selector records this equality, and the
third permutation check proves that the selected rows of $\widetilde D_e$
form exactly $D_{out}$.

Overall, ZKGraph's set-based expansion construction requires three permutation
checks, one lookup check, and four range checks. Because each range check
performs multiple limb lookups per row, these four range checks constitute the
main bottleneck in the proof generation phase.

\subsubsection*{Our circuit design}
Our construction reduces this proving overhead by avoiding edge-table sorting
and per-edge bracketing, cutting the required range checks from four to one.
It proves that $D_{out}$ contains all
and only the input edges whose source identifiers belong to $ID_s$.
We decompose this target into three conditions: the output and non-output
edges together form the input; every output source belongs to $ID_s$; and
no non-output source belongs to $ID_s$. To certify the last condition, we
prove that the distinct source identifiers of the non-output edges are
disjoint from $ID_s$. The prover first constructs the following witnesses:
\begingroup
\setlength{\leftmargini}{1.2em}
\begin{itemize}
\item $D_e^n=(A^n,B^n)$, the non-output edge table consisting of the
rows of $D_e$ whose source identifiers do not belong to $ID_s$, where
$A^n$ stores their source identifiers and $B^n$ stores their target identifiers;
\item $A^{n,d}$, a duplicate-free column containing one representative
of each distinct source identifier appearing in $A^n$; this column is
used to prove collectively that the sources of all non-output edges lie
outside $ID_s$;
\item $ID_{union}$, a sorted column obtained from the concatenation of
$A^{n,d}$ and $ID_s$; its strict ordering ensures that no
identifier appears in both columns.
\end{itemize}
\endgroup

The prover then proves the following constraints. First, a permutation
check proves that concatenating $D_{out}$ and $D_e^n$ yields a
permutation of the input edge table $D_e$. Using a random challenge
$\alpha$, we compress the rows as $C_{out}[i] = C[i]+\alpha D[i]$,
$C_n[i] = A^n[i]+\alpha B^n[i]$, and $C_e[i] = A[i]+\alpha B[i]$.
With another random challenge $\beta$, the prover proves
\[
  \sum_{i=0}^{|C|-1} \frac{1}{\beta+C_{out}[i]}
  +
  \sum_{i=0}^{|A^n|-1} \frac{1}{\beta+C_n[i]}
  =
  \sum_{i=0}^{|A|-1} \frac{1}{\beta+C_e[i]}.
\]
This constraint ensures that every row occurrence in $D_e$ is accounted for
exactly once by either $D_{out}$ or $D_e^n$, and that neither table
contains a fabricated edge row.

Second, a lookup check proves that every source identifier in $A^n$
appears in $A^{n,d}$. This ensures that $A^{n,d}$ covers all source
identifiers represented in the non-output edge table.

Third, a permutation check proves that concatenating $A^{n,d}$ and
$ID_s$ yields a permutation of $ID_{union}$. A range check further proves
$ID_{union}[i] < ID_{union}[i+1]$ for every adjacent pair. Any identifier
shared by $A^{n,d}$ and $ID_s$ would appear twice in $ID_{union}$, violating
its strict ordering. These checks therefore establish that the two
columns share no identifier.

Fourth, a lookup check proves that every source identifier in the
output column $C$ belongs to $ID_s$, so every
reported edge satisfies the set-based expansion condition.

Compared with ZKGraph, our construction keeps the same number of linear-size
permutation and lookup checks (two permutation checks and two lookup checks,
versus three permutation checks and one lookup check in ZKGraph) while
substantially reducing the range-check workload. Specifically, ZKGraph requires
four range checks: three over $|A|$-row edge columns to sort $\widetilde{A}$
($\widetilde{A}[i]\leq\widetilde{A}[i+1]$) and bracket every edge source
($L[i]\leq\widetilde{A}[i]$ and $\widetilde{A}[i]<U[i]$), plus one over the
$(|ID_s|+2)$-row column $\widetilde{ID}_s$
($\widetilde{ID}_s[i]<\widetilde{ID}_s[i+1]$). In contrast, our construction
eliminates edge-table sorting and per-edge bracketing, requiring only a single
range check over $ID_{union}$ ($|A^{n,d}|+|ID_s|$ rows) to prove
$ID_{union}[i]<ID_{union}[i+1]$. Even in the worst case where all non-output
edges have distinct sources ($|A^{n,d}|\leq |A|$), this replaces three
$|A|$-row range checks and one $(|ID_s|+2)$-row range check with a single range
check of at most $|A|+|ID_s|$ rows. Moreover, when multiple non-output edges
share a source vertex, deduplication makes $|A^{n,d}|$ smaller than $|A|$,
further shrinking the range-checked column to distinct identifiers.

Let $E$ denote the number of input edges, $S$ denote the size of
$ID_s$, $T$ denote the size of the range table, and $n$ denote the
number of limb lookup checks per range check.
Since the output and non-output edge tables together contain $E$ rows,
and $A^{n,d}$ contains at most $E$ identifiers, the two permutation
checks have total circuit complexity $O(E+S)$, as do the two lookup checks.
The range check establishes strict ordering over $ID_{union}$, which
contains at most $E+S$ identifiers, with circuit complexity $O(n(E+S+T))$.
Together, these constraints yield an overall circuit complexity
of $O(n(E+S+T))$.

\subsubsection*{Correctness}
The correctness follows from the following facts. The first-step
permutation check establishes that, as multisets,
$D_{out}$ and $D_e^n$ together contain exactly the rows of $D_e$; hence,
every input-edge occurrence is assigned to one of these two tables,
with no row inserted or omitted. The second-step lookup check proves
that $A^{n,d}$ covers every source identifier in $A^n$. The third-step
permutation and range checks establish that $A^{n,d}$ and $ID_s$ share
no identifier. Together, the second and third steps therefore prove
that no edge in $D_e^n$ starts from a vertex in $ID_s$. Finally, the
fourth-step lookup check proves that every source identifier in $C$
belongs to $ID_s$, so every edge in $D_{out}$ satisfies the expansion
condition. These constraints jointly prove that $D_{out}$ contains all and
only the one-hop traversals that start from vertices in $ID_s$.

\subsubsection*{Bidirectional relations}
In our benchmark queries, the query plans of both ZKGraph and VerfGraph require
set-based expansion only over directed edge relations. Since the bidirectional
extension also has potential applications, the directed construction can be
extended to a bidirectional relation $D_e=(A,B)$ by applying it to the two views
$(A,B)$ and $(B,A)$, combining their outputs as a multiset, and counting
self-loops once. Implementing this extension and evaluating its
proving overhead are left to future work.

\subsection{Single-Source Shortest Path}
\label{sec:sssp}

Given a source identifier $id_s$, an edge table $D_e=(A,B)$, and a column $N$
containing all distinct node identifiers, SSSP returns a distance column $D$
aligned with $N$, where $D[i]$ records the shortest-path distance from $id_s$
to $N[i]$. If $N[i]$ is unreachable from $id_s$, $D[i]$ is set to a predefined
sentinel $d_{\max}$ larger than any finite shortest-path distance (for $h$-hop
queries, setting $d_{\max}=h+1$ likewise marks nodes beyond $h$ hops as
unreachable).\footnote{Edge weights, simple-path lengths, and $d_{\max}$ are bounded
nonnegative integers well below the field modulus (and the range-check bound
when applicable), avoiding modular wrap-around.} We first consider unweighted
SSSP, where every edge has unit cost, and then extend the construction to
weighted SSSP with positive integer edge weights.

\subsubsection*{Unweighted SSSP}
Instead of unrolling multi-hop traversals inside the circuit, ZKGraph verifies
directed unweighted SSSP by combining a node-level upper-bound check with an
edge-level lower-bound check, a framework that VerfGraph retains for directed
unweighted relations. On the node side, the prover computes shortest paths
locally and supplies a Boolean source flag $F_s$, a predecessor column $P$, and
a predecessor-distance column $D_P$ aligned with $N$. Polynomial constraints
enforce $F_s[i]=1$ iff $N[i]=id_s$ and require $D[i]=0$ on this source row.
Every non-source row $i$ ($F_s[i]=0$) is constrained to satisfy either
$D[i]=d_{\max}$ when $N[i]$ is unreachable or $D[i]=D_P[i]+1$ (with
$D_P[i]\neq d_{\max}$) when $N[i]$ is
reachable. In the reachable case, a lookup check proves $(P[i],D_P[i])\in(N,D)$
so that $D_P[i]$ matches the distance recorded for $P[i]$, and a second lookup
check proves $(P[i],N[i])\in(A,B)$ so that the step follows an actual input
edge. Together, these node-level checks establish an upper bound by exhibiting a
valid path from $id_s$ of length $D[i]$ for each node with $D[i]\neq d_{\max}$.

Node-level checks alone, however, do not prevent the prover from reporting a
longer detour or marking a reachable node as $d_{\max}$. To rule out both cases,
the circuit enforces edge relaxation across all rows of $D_e=(A,B)$.
Because the circuit cannot dynamically index $(N,D)$ by endpoint identifier,
the prover supplies two edge-aligned columns $D_A$ and $D_B$ that record the
distances reported in $D$ for endpoints $A[j]$ and $B[j]$ of each edge row $j$.
Two lookup checks bind $(A[j],D_A[j])$ and $(B[j],D_B[j])$ to $(N,D)$, and a
range check enforces $D_B[j]\leq D_A[j]+1$, ensuring that no edge admits a
shorter path than the reported distances.

For a bidirectional edge relation, ZKGraph's proof construction uses BiRC to
place the smaller identifier first in each input edge and each predecessor pair
$(P[i],N[i])$. The circuit checks that reordering preserves the endpoints and
uses two range checks, one for each table, to enforce their order. A lookup
check then proves that each reordered pair appears in the reordered input edge
table. The remaining distance checks reuse the directed construction and
enforce $D_B[j]\leq D_A[j]+1$ in the stored direction, whereas bidirectional
optimality also requires $D_A[j]\leq D_B[j]+1$.

VerfGraph eliminates all range checks in bidirectional unweighted SSSP while
enforcing relaxation in both directions. First, following
Section~\ref{sec:single-source-expansion}, we verify predecessor edges against
the two directed views $(A,B)$ and $(B,A)$ without duplicating edge rows or
reordering endpoints. Let $R$ denote the row indices of non-source nodes
with $D[i]\neq d_{\max}$. Using a random challenge $\alpha$, we compress the
predecessor pairs and the two edge views as $C_{pred}[i]=P[i]+\alpha N[i]$,
$C_{input}[j]=A[j]+\alpha B[j]$, and $C'_{input}[j]=B[j]+\alpha A[j]$. Following
the lookup protocol in Section~\ref{sec:prelim-checks}, the prover supplies
multiplicity columns $m_1$ and $m_2$ for forward and reverse matches, and with a
second challenge $\beta$ proves
\begingroup
\thinmuskip=2mu \medmuskip=2mu \thickmuskip=3mu
\[
  \sum_{i\in R}\!\frac{1}{\beta+C_{pred}[i]}
  =
  \sum_{j=0}^{|A|-1}\!\left(
    \frac{m_1[j]}{\beta+C_{input}[j]}
    +\frac{m_2[j]}{\beta+C'_{input}[j]}
  \right).
\]
\endgroup
Second, over a bidirectional edge, either both endpoints are reachable with
shortest distances differing by at most $1$, or both lie in an unreachable
component and take $d_{\max}$. Enforcing both $D_B[j]\leq D_A[j]+1$ and
$D_A[j]\leq D_B[j]+1$ is therefore equivalent to requiring the integer
difference $D_A[j]-D_B[j]$ to lie in $\{-1,0,1\}$. Rather than using range
checks, we enforce this three-value condition directly with the polynomial
constraint
\[
  (D_A[j]-D_B[j])\bigl((D_A[j]-D_B[j])^2-1\bigr)=0,
\]
which avoids both BiRC's range checks and edge-relaxation range checks.

\subsubsection*{Weighted SSSP}
To support weighted graphs, we extend $D_e$ with an aligned column $W$ of
positive integer edge weights, forming $(A,B,W)$. On the node side, the prover
supplies an additional predecessor-weight column $W_P$, where $W_P[i]$ records
the weight of the edge connecting $P[i]$ to $N[i]$. Because edge costs are no
longer uniform, verifying edge existence alone is insufficient: for each
reachable non-source node, the predecessor-edge lookup check must jointly bind
the triple $(P[i],N[i],W_P[i])$ to $(A,B,W)$ in the directed case, or to
$(A,B,W)$ and $(B,A,W)$ via the same two-view lookup check in the bidirectional
case. The distance-propagation constraint then becomes $D[i]=D_P[i]+W_P[i]$.

On the edge side, two lookup checks still bind $(A[j],D_A[j])$ and
$(B[j],D_B[j])$ to $(N,D)$, while relaxation incorporates $W[j]$. For directed
relations, one range check enforces $D_B[j]\leq D_A[j]+W[j]$. For bidirectional
relations, two range checks enforce relaxation in both directions,
$D_B[j]\leq D_A[j]+W[j]$ and $D_A[j]\leq D_B[j]+W[j]$, while still avoiding
BiRC's range checks.

We now summarize the circuit complexity across all four SSSP variants. Let $E$
denote the number of input edges, $V$ the number of nodes, $T$ the range-table
size, and $n$ the number of limb lookup checks per range check. In every
variant, node-level polynomial constraints and the predecessor-distance lookup
check have circuit complexity $O(V)$, while the predecessor-edge lookup check
and the two endpoint-distance lookup checks have circuit complexity $O(E+V)$.
The variants differ in their edge-relaxation checks. Directed unweighted SSSP
uses one $E$-row range check of complexity $O(n(E+T))$, yielding an overall
circuit complexity of $O(V+n(E+T))$. Bidirectional unweighted SSSP replaces
relaxation range checks with $O(E)$ polynomial constraints and avoids BiRC,
reducing the overall circuit complexity to $O(E+V)$, independent of both $n$
and $T$. Finally, directed and bidirectional weighted SSSP require one and two
$E$-row range checks, respectively, giving $O(V+n(E+T))$ in both cases.

\subsubsection*{Correctness}
Correctness follows from combining the node-level upper bound with the
edge-level lower bound. First, because $N$ lists all distinct nodes, $F_s$
uniquely identifies $id_s$ and fixes its distance to $0$. For every non-source
node with $D[i]\neq d_{\max}$, the node-level lookups and constraints guarantee
that $P[i]$ appears in $(N,D)$ with distance $D_P[i]\neq d_{\max}$, connects to $N[i]$ via a
valid input edge (with verified weight $W_P[i]$ in the weighted case), and
satisfies $D[i]$ equal to $D_P[i]$ plus that edge's cost. Because edge costs are
positive, tracing predecessors backward from $N[i]$ strictly decreases distance
and must terminate at $id_s$, establishing a valid path from $id_s$ to $N[i]$ of
total cost $D[i]$.

Second, the endpoint-distance lookups ensure that edge relaxation is evaluated
on these exact node distances from $(N,D)$. Applying the relaxation inequality
inductively along any valid path from $id_s$ to $N[i]$ (in either direction for
bidirectional relations) implies that $D[i]$ cannot exceed that path's total
cost, so no shorter path from $id_s$ to $N[i]$ can exist. Combined with the
valid path of cost $D[i]$ above, every entry $D[i]\neq d_{\max}$ equals the true
shortest-path distance. Finally, $D[i]=d_{\max}$ holds iff $N[i]$ is
unreachable: if $N[i]$ is reachable, the inductive bound along a shortest path
forces $D[i]$ below $d_{\max}$, whereas if $N[i]$ is unreachable, no valid path
from $id_s$ exists to support $D[i]\neq d_{\max}$.

\begin{table*}[!t]
  \caption{Circuit complexity of operator-proof constructions.}
  \label{tab:operator-cost-comparison}
  \centering
  \footnotesize
  \setlength{\tabcolsep}{2.4pt}
  \setlength{\aboverulesep}{0.25ex}
  \setlength{\belowrulesep}{0.38ex}
  \renewcommand{\arraystretch}{1.04}
  \begin{tabular*}{\textwidth}{@{\extracolsep{\fill}}cccccc@{}}
    \toprule
    Operator & VerfGraph & ZKGraph & Operator & VerfGraph & ZKGraph \\
    \midrule
    Directed single-source expansion
      & $O(E)$ & $O(E)$
      & Bidirectional single-source expansion
      & $O(E)$ & $O(n(E+T))$ \\
    Directed unweighted SSSP
      & $O(V+n(E+T))$ & $O(V+n(E+T))$
      & Bidirectional unweighted SSSP
      & $O(E+V)$ & $O(n(E+V+T))$ \\
    Directed weighted SSSP
      & $O(V+n(E+T))$ & ---
      & Bidirectional weighted SSSP
      & $O(V+n(E+T))$ & --- \\
    Set-based expansion
      & $O(n(E+S+T))$ & $O(n(E+S+T))$
      & Count & $O(I+n(G+T))$ & --- \\
    \bottomrule
  \end{tabular*}
  \par\vspace{2pt}
  \begin{tabular}{@{}p{0.48\textwidth}@{\hspace{0.04\textwidth}}p{0.48\textwidth}@{}}
    $E$: number of input edges in expansion and SSSP.
      & $n$: number of limb lookup checks per range check. \\
    $S$: number of vertices in the source set $ID_s$ of set-based expansion.
      & $V$: number of vertices in the graph processed by SSSP. \\
    $I$: number of input values processed by count.
      & $G$: number of output groups produced by count. \\
    $T$: size of the range table used for range checks.
      & ---: ZKGraph does not provide a corresponding operator-proof construction.
  \end{tabular}
\end{table*}

\subsection{Count}
\label{sec:count}

The count operator groups identical input values and returns the number of
occurrences of each distinct value, supporting common graph-query tasks such
as computing vertex degrees, tag frequencies, and per-type edge counts.
Given an input vector $A$ (with $|A|$ below the field modulus), it outputs
$D_{out}=(B,C)$, where $B$ contains the distinct values in $A$ and $C[i]$
records the number of occurrences of $B[i]$ in $A$.

\subsubsection*{Circuit design}
With a random challenge $\beta$, the prover first checks that the input
multiset $A$ matches the weighted output multiset $(B,C)$:
\[
  \sum_{i=0}^{|A|-1} \frac{1}{\beta+A[i]}
  =
  \sum_{i=0}^{|B|-1} \frac{C[i]}{\beta+B[i]}.
\]
To exclude zero-count keys absent from $A$, the prover also supplies an
inverse column $C_{\mathrm{inv}}$ and enforces $C[i]\cdot C_{\mathrm{inv}}[i]=1$
on every output row. A range check then enforces $B[i+1]>B[i]$ for each
adjacent pair so that $B$ contains no duplicate keys. Requiring $B$ to appear
in ascending order does not affect the semantics of count.

Let $I$ denote the number of input values, $G$ the number of output groups,
$T$ the range-table size, and $n$ the number of limb lookup checks per range
check. The weighted multiset check and nonzero-count constraints have circuit
complexity $O(I+G)$, and the range check over the $G$ output keys has circuit
complexity $O(n(G+T))$, yielding an overall circuit complexity of
$O(I+n(G+T))$.

\subsubsection*{Correctness}
The range check prevents the prover from splitting the occurrences of a value
in $A$ across multiple output rows, while $C[i]\cdot C_{\mathrm{inv}}[i]=1$
forces $C[i]\neq 0$ and rules out keys absent from $A$. Together with the
weighted multiset equality (and $|A|$ below the field modulus), every distinct
value in $A$ appears once in $B$ with its exact count $C[i]$. Thus, $(B,C)$
correctly encodes the group counts of $A$.

\subsection{Circuit Complexity Comparison}
\label{sec:cost-comparison}

Table~\ref{tab:operator-cost-comparison} compares the circuit complexity of
VerfGraph and ZKGraph. For bidirectional single-source expansion and
bidirectional unweighted SSSP, VerfGraph eliminates range checks,
making its circuit complexity independent of both the limb count $n$ and the
range-table size $T$, thereby reducing proof-generation cost. For set-based expansion, although
VerfGraph and ZKGraph have the same asymptotic circuit complexity, VerfGraph's
proof construction reduces costly range checks from four to one,
yielding a constant-factor reduction in proof-generation cost.

\subsection{Summary}
\label{sec:operator-summary}

The four operators presented above provide core building blocks for verifying
graph queries under the expansion-centric model. Under this model, other
operations include filtering, order-by and limit-$k$ (top-$k$), reachability,
and all-shortest-path enumeration. For these operations, VerfGraph retains
ZKGraph's operator-level proof strategies (detailed in ZKGraph's arXiv
preprint and public implementation). Together, the operators needed for a
query are composed into a query plan, passing intermediate relations from
upstream operators to downstream ones.

To verify this composed data flow, opening proofs against the database
commitment $C_G$ confirm that base-table inputs match the committed columns
(Section~\ref{sec:security-model}), while AIR constraints enforce each
operator's input--output relationship. Across connected operators in the
query plan, these constraints further require each consumer's input to match
the corresponding producer's output and bind the final output to the returned
answer. VerfGraph thus generates a single zero-knowledge proof for the
complete query evaluation, allowing the verifier to check the answer without
accessing private graph records or intermediate relations.

\section{Experiments}
\label{sec:experiments}

\subsection{Experimental Setup}
\label{sec:experimental-setup}

We implement VerfGraph in Rust on top of Plonky3, which we choose for
its AIR arithmetization framework and FRI-based PCS.
This PCS is hash-based, allowing the use of small prime fields for
more efficient proof arithmetic. For this reason, we choose Goldilocks64,
a widely used 64-bit prime field that fits a single machine word and
supports fast FFT-based polynomial operations for proof generation.
We configure both systems to operate at a comparable security level:
ZKGraph uses the KZG PCS instantiated over the BN254 elliptic curve,
which is estimated to provide about 100 bits of classical
security~\cite{barbulescu2019keysize}. We accordingly configure our
FRI-based PCS and quadratic extension-field challenges to target the same security level.

We use a subset of the LDBC Social Network Benchmark (SNB)
Interactive workload~\cite{ldbcSNBDocs} as our query workload:
IS3, IS4, IS5, IC1, IC2, IC9,\footnote{ZKGraph's
paper refers to this query as IC8, while its public code repository and query
semantics correspond to IC9. We follow the latter.} and IC13.
These seven queries are exactly the queries used in ZKGraph's
experiments, which enables a direct comparison with ZKGraph.
Beyond ZKGraph's query coverage, we further include IC6 and IC14 V2.
IC6 relies on our count proof construction, and IC14 V2 relies
on our weighted SSSP proof construction.

Following ZKGraph, we measure the database scale of each query by the number
of rows in its relevant fact table.
The relevant fact table is
\texttt{person\_\allowbreak{}knows\_\allowbreak{}person}
for IS3, IC1, and IC13;
\texttt{comment\_\allowbreak{}hasCreator\_\allowbreak{}person}
for IS4, IS5, IC2, and IC9; and
\texttt{post\_\allowbreak{}hasTag\_\allowbreak{}tag} for IC6.
For IC14 V2, we use the combined number of rows in
\texttt{comment\_\allowbreak{}replyOf\_\allowbreak{}comment} and
\texttt{comment\_\allowbreak{}replyOf\_\allowbreak{}post}.
Unless otherwise stated, we set this row count to
60k, 120k, and 240k to examine how performance scales with
database size. These scales match those in ZKGraph's experiments. For the seven shared
queries, both systems use the 60k and 120k datasets from ZKGraph's public
code repository, along with a 240k dataset constructed with the same
relation schemas and table organization (as the repository does not provide
a 240k dataset).

All experiments were conducted on a server with 24 vCPUs
(Intel Ice Lake, 3.0 GHz) and 192 GiB of RAM, running 64-bit
Ubuntu 24.04. All reported results are averaged over multiple runs.

\subsection{System-Level Performance}
\label{sec:benchmark-zkgraph}

This subsection evaluates VerfGraph's offline database-commitment and
circuit-preprocessing costs, its system-level performance compared with
ZKGraph, and additional queries enabled by our count
and weighted SSSP constructions.

\subsubsection*{Database Commitment and Circuit Preprocessing}
We measure database commitment time and circuit preprocessing time.
For each query template, the database commitment covers
the input tables and columns used by that query.
The experimental results are shown in
Tables~\ref{tab:database-commitment-time}
and~\ref{tab:circuit-preprocessing-time}.

\begin{table}[!t]
  \caption{Database commitment time by query and database scale (s).}
  \label{tab:database-commitment-time}
  \centering
  \footnotesize
  \setlength{\tabcolsep}{1.8pt}
  \setlength{\aboverulesep}{0.25ex}
  \setlength{\belowrulesep}{0.35ex}
  \renewcommand{\arraystretch}{1.02}
  \begin{tabular}{@{}cccccccccc@{}}
    \toprule
    Scale & IS3 & IS4 & IS5 & IC1 & IC2 & IC6 & IC9 & IC13 & IC14 V2 \\
    \midrule
    60k & 0.053 & 0.057 & 0.064 & 0.163 & 0.096 & 0.065 & 0.097 & 0.045 & 0.145 \\
    120k & 0.105 & 0.109 & 0.110 & 0.306 & 0.196 & 0.136 & 0.195 & 0.088 & 0.285 \\
    240k & 0.208 & 0.219 & 0.220 & 0.596 & 0.396 & 0.279 & 0.396 & 0.173 & 0.570 \\
    \bottomrule
  \end{tabular}
\end{table}

\begin{table}[!t]
  \caption{Circuit preprocessing time by query and database scale (s).}
  \label{tab:circuit-preprocessing-time}
  \centering
  \footnotesize
  \setlength{\tabcolsep}{1.8pt}
  \setlength{\aboverulesep}{0.25ex}
  \setlength{\belowrulesep}{0.35ex}
  \renewcommand{\arraystretch}{1.02}
  \begin{tabular}{@{}cccccccccc@{}}
    \toprule
    Scale & IS3 & IS4 & IS5 & IC1 & IC2 & IC6 & IC9 & IC13 & IC14 V2 \\
    \midrule
    60k & 0.043 & 0.041 & 0.042 & 0.271 & 0.043 & 0.042 & 0.043 & 0.089 & 0.300 \\
    120k & 0.083 & 0.081 & 0.080 & 0.520 & 0.084 & 0.082 & 0.085 & 0.170 & 0.603 \\
    240k & 0.163 & 0.156 & 0.156 & 1.041 & 0.164 & 0.163 & 0.164 & 0.330 & 1.222 \\
    \bottomrule
  \end{tabular}
\end{table}

\subsubsection*{Benchmarking with ZKGraph}
ZKGraph is the state-of-the-art zero-knowledge verifiable graph
query evaluation system~\cite{wu2025zkgraph}.
We compare VerfGraph with ZKGraph on the seven queries evaluated in
ZKGraph's experiments (IS3, IS4, IS5, IC1, IC2, IC9, and IC13) at the
60k, 120k, and 240k fact-table scales.
Both systems use the same datasets,
and VerfGraph adopts ZKGraph's simplified value-encoding rules for nonnumeric values
such as strings and dates, so that both systems prove the same query results
over the same encoded data.

For each query, we use ZKGraph's operator decomposition, replacing
its operator proofs with the circuits in Section~\ref{sec:circuit-design}.
Following ZKGraph, we prove friends-of-friends queries with bidirectional
SSSP instead of sequential expansions.

We compare proof-generation time, peak prover memory usage,
proof size, and verification time. The experimental results
are shown in Table~\ref{tab:zkgraph-comparison}.

\begin{table*}[!t]
  \caption{Query performance of VerfGraph, ZKGraph, and VG-Halo2.}
  \label{tab:zkgraph-comparison}
  \centering
  \begin{minipage}{0.94\textwidth}
  \footnotesize
  \setlength{\tabcolsep}{3.0pt}
  \setlength{\aboverulesep}{0.22ex}
  \setlength{\belowrulesep}{0.32ex}
  \renewcommand{\arraystretch}{1.0}
  \begin{tabular*}{\textwidth}{@{\extracolsep{\fill}}cccccccccccccc@{}}
\toprule
\multirow{2}{*}{Query} & \multirow{2}{*}{Scale}
  & \multicolumn{3}{c}{\shortstack{Proof Generation\\Time (s)}}
  & \multicolumn{3}{c}{Proof Size (KiB)}
  & \multicolumn{3}{c}{\shortstack{Verification\\Time (ms)}}
  & \multicolumn{3}{c}{\shortstack{Peak Memory\\Usage (GiB)}} \\
\cmidrule(lr){3-5}\cmidrule(lr){6-8}\cmidrule(lr){9-11}\cmidrule(lr){12-14}
  & & {\scriptsize VerfGraph} & {\scriptsize ZKGraph} & {\scriptsize VG-Halo2}
  & {\scriptsize VerfGraph} & {\scriptsize ZKGraph} & {\scriptsize VG-Halo2}
  & {\scriptsize VerfGraph} & {\scriptsize ZKGraph} & {\scriptsize VG-Halo2}
  & {\scriptsize VerfGraph} & {\scriptsize ZKGraph} & {\scriptsize VG-Halo2} \\
\midrule
\multirow{3}{*}{IC1} & 60k & 1.742 & 77.880 & 43.374 & 679.079 & 57.938 & 32.406 & 24.904 & 21.351 & 14.330 & 1.678 & 7.947 & 5.822 \\
 & 120k & 3.409 & 156.440 & 87.175 & 718.237 & 57.938 & 32.406 & 27.240 & 21.435 & 14.388 & 3.152 & 15.901 & 11.656 \\
 & 240k & 6.681 & 330.215 & 181.695 & 758.958 & 57.938 & 32.406 & 28.558 & 21.581 & 32.889 & 6.095 & 31.380 & 22.990 \\
\midrule
\multirow{3}{*}{IC2} & 60k & 1.050 & 56.143 & 30.368 & 558.924 & 40.250 & 23.094 & 18.986 & 16.229 & 10.779 & 0.941 & 5.752 & 3.882 \\
 & 120k & 2.071 & 113.194 & 61.477 & 598.082 & 40.250 & 23.094 & 20.296 & 16.372 & 10.823 & 1.789 & 11.511 & 7.761 \\
 & 240k & 4.114 & 240.029 & 129.966 & 638.803 & 40.250 & 23.094 & 21.616 & 16.207 & 10.738 & 3.512 & 23.176 & 15.261 \\
\midrule
\multirow{3}{*}{IC9} & 60k & 1.099 & 60.893 & 32.677 & 566.096 & 43.688 & 24.500 & 19.307 & 17.170 & 11.802 & 1.001 & 6.069 & 4.028 \\
 & 120k & 2.164 & 122.647 & 65.952 & 605.254 & 43.688 & 24.500 & 20.558 & 17.326 & 11.714 & 1.886 & 12.135 & 8.213 \\
 & 240k & 4.294 & 256.487 & 138.687 & 645.975 & 43.688 & 24.500 & 21.907 & 17.246 & 12.970 & 3.662 & 24.051 & 15.889 \\
\midrule
\multirow{3}{*}{IC13} & 60k & 0.402 & 13.622 & 5.191 & 460.734 & 9.000 & 3.406 & 14.872 & 6.745 & 4.918 & 0.327 & 1.481 & 0.790 \\
 & 120k & 0.768 & 27.195 & 10.403 & 499.893 & 9.000 & 3.406 & 16.137 & 6.842 & 4.870 & 0.630 & 2.958 & 1.514 \\
 & 240k & 1.525 & 56.903 & 21.339 & 540.613 & 9.000 & 3.406 & 17.367 & 6.667 & 4.964 & 1.239 & 5.860 & 3.038 \\
\midrule
\multirow{3}{*}{IS3} & 60k & 0.630 & 13.975 & 8.332 & 499.278 & 10.219 & 6.000 & 16.358 & 7.178 & 5.912 & 0.536 & 1.612 & 1.027 \\
 & 120k & 1.254 & 28.652 & 16.868 & 539.702 & 10.219 & 6.000 & 17.671 & 7.070 & 5.893 & 1.046 & 3.275 & 2.087 \\
 & 240k & 2.628 & 60.454 & 35.020 & 585.907 & 10.219 & 6.000 & 19.273 & 7.266 & 6.091 & 2.182 & 6.561 & 4.187 \\
\midrule
\multirow{3}{*}{IS4} & 60k & 0.336 & 3.899 & 2.205 & 451.822 & 2.562 & 1.844 & 14.548 & 4.569 & 4.020 & 0.295 & 0.793 & 0.376 \\
 & 120k & 0.648 & 7.871 & 4.354 & 490.980 & 2.562 & 1.844 & 15.773 & 4.559 & 3.969 & 0.578 & 1.600 & 0.781 \\
 & 240k & 1.269 & 16.341 & 8.669 & 531.701 & 2.562 & 1.844 & 17.017 & 4.727 & 3.942 & 1.146 & 3.382 & 1.574 \\
\midrule
\multirow{3}{*}{IS5} & 60k & 0.379 & 4.015 & 3.234 & 453.088 & 2.781 & 2.062 & 14.554 & 4.466 & 4.243 & 0.292 & 0.812 & 0.626 \\
 & 120k & 0.670 & 8.111 & 6.480 & 492.246 & 2.781 & 2.062 & 15.866 & 4.387 & 4.181 & 0.573 & 1.635 & 1.229 \\
 & 240k & 1.331 & 16.721 & 13.349 & 532.967 & 2.781 & 2.062 & 17.102 & 4.500 & 4.365 & 1.132 & 3.348 & 2.573 \\
\bottomrule
\end{tabular*}

  \end{minipage}
\end{table*}

\textbf{Proof-generation time.}
VerfGraph substantially reduces proof-generation time over ZKGraph.
The speedup is most pronounced on complex IC queries,
reaching $33.9\times$--$59.7\times$.\footnote{The public ZKGraph implementation commits
its database witness columns during proof generation. Even when VerfGraph's
database-commitment time is included in its proof-generation time, VerfGraph
remains $33.5\times$--$54.7\times$ faster across the four IC queries at the
240k scale.}

\textbf{Peak prover memory usage.}
VerfGraph also substantially reduces peak prover memory usage over ZKGraph;
the reduction is again more significant on complex IC queries
($4.5\times$--$6.6\times$).

\textbf{Proof size.}
The proof size of VerfGraph is larger than that of ZKGraph,
but remains sub-MB for every query and scale.
This increase reflects our use of an FRI-based PCS
and is an acceptable tradeoff for the substantially faster proof generation
and lower memory usage reported above.

\textbf{Verification time.}
VerfGraph maintains millisecond-level verification across all queries
and scales.

Across the three scales, as database scale increases, proof-generation time
grows accordingly, while proof size and verification time grow only slowly.

\subsubsection*{Additional Query Coverage}
\label{sec:additional-query-coverage}

Beyond the seven queries shared with ZKGraph, we evaluate IC6
and IC14 V2 at the 60k, 120k, and 240k scales.
IC6 uses our count proof construction, and IC14 V2 uses our
weighted SSSP proof construction. The experimental results are
shown in Table~\ref{tab:additional-queries}.

\begin{table}[!t]
  \caption{Performance of additional LDBC SNB queries in
  VerfGraph.}
  \label{tab:additional-queries}
  \centering
  \footnotesize
  \setlength{\tabcolsep}{2.0pt}
  \setlength{\aboverulesep}{0.25ex}
  \setlength{\belowrulesep}{0.35ex}
  \renewcommand{\arraystretch}{1.02}
  \begin{tabular}{@{}cccccc@{}}
\toprule
\multirow{2}{*}{Query} & \multirow{2}{*}{Scale}
  & \shortstack{Proof\\Generation} & Proof Size & Verification & Peak Memory \\
  & & Time (s) & (KiB) & Time (ms) & Usage (GiB) \\
\midrule
\multirow{3}{*}{IC6} & 60k & 2.502 & 566.394 & 18.908 & 1.852 \\
 & 120k & 4.900 & 605.552 & 20.184 & 3.755 \\
 & 240k & 9.877 & 646.272 & 21.502 & 7.441 \\
\midrule
\multirow{3}{*}{IC14 V2} & 60k & 2.028 & 593.514 & 19.795 & 1.811 \\
 & 120k & 4.035 & 634.234 & 21.180 & 3.546 \\
 & 240k & 8.201 & 676.518 & 22.482 & 7.025 \\
\bottomrule
\end{tabular}

\end{table}

Thus, VerfGraph supports both queries
while retaining sub-MB proofs and millisecond-level verification
at all scales.

\subsection{Operator Proof-Generation Time}
\label{sec:operation-performance}

To evaluate the proving costs of the operator-proof constructions introduced
in Section~\ref{sec:circuit-design}, we measure single-source expansion and
the two set-based expansions in IC2; unweighted SSSP and
count in IC6; and weighted SSSP in IC14 V2. These operators cover all
constructions introduced in Section~\ref{sec:circuit-design}.

We construct a standalone circuit for each operator and measure its
proof-generation time using inputs from the corresponding query at the
60k database scale; the two set-based expansions in IC2 are proved jointly
in one circuit. The experimental results are shown in
Table~\ref{tab:operator-performance}.

\begin{table}[!t]
  \caption{Standalone operator-proof-generation time at the 60k scale.}
  \label{tab:operator-performance}
  \centering
  \footnotesize
  \setlength{\tabcolsep}{4pt}
  \setlength{\aboverulesep}{0.25ex}
  \setlength{\belowrulesep}{0.35ex}
  \renewcommand{\arraystretch}{1.02}
  \begin{tabular}{@{}ccc@{}}
    \toprule
    \multirow{2}{*}{Operator} & \multirow{2}{*}{Source query} & Proof generation \\
     & & time (s) \\
    \midrule
    Single-source expansion & IC2 & 0.315 \\
    Set-based expansion $\times 2$ & IC2 & 0.484 \\
    Unweighted SSSP & IC6 & 0.666 \\
    Count & IC6 & 0.336 \\
    Weighted SSSP & IC14 V2 & 0.472 \\
    \bottomrule
  \end{tabular}
\end{table}

\subsection{Same-Backend Comparison with ZKGraph}
\label{sec:protocol-performance}

To show that the improvements in Section~\ref{sec:benchmark-zkgraph} come
not only from the change in proving backend but also from our proposed
operator-proof constructions, we implement our constructions for a
same-backend comparison. This implementation, referred to as VG-Halo2,
retains the Halo2/KZG backend, ZKGraph's operator decomposition, and the
mechanisms for permutation, lookup, and 8-bit limb range checks, and
replaces ZKGraph's operator-proof constructions with ours. In ZKGraph's public implementation,
the database commitment is computed during proof generation. For
consistency with ZKGraph, VG-Halo2 also performs the database commitment in
this phase.

\subsubsection*{System-Level Comparison}
We compare VG-Halo2 with ZKGraph on the seven queries and three database
scales used in Section~\ref{sec:benchmark-zkgraph}. The experimental results
are shown in the ZKGraph and VG-Halo2 columns of
Table~\ref{tab:zkgraph-comparison}. Compared with ZKGraph, VG-Halo2 reduces
proof-generation time, proof size, and peak prover memory usage on all seven
queries, especially on the complex IC
queries. For these queries, VG-Halo2 reduces proof-generation time by
$44.3\%$--$62.5\%$, proof size by $42.6\%$--$62.2\%$, and peak prover memory
usage by $26.7\%$--$48.8\%$, while maintaining millisecond-level
verification.

Beyond these four metrics, we also compare the circuit-specific
key-generation time of the two systems, which serves as the preprocessing
cost for the Halo2-based implementations. Although it plays the same role as
circuit preprocessing in Section~\ref{sec:benchmark-zkgraph}, the underlying
preprocessing operations differ between Halo2 and Plonky3, so we compare
key-generation time directly between VG-Halo2 and ZKGraph. The experimental
results are shown in Table~\ref{tab:keygen-comparison}, and the two systems
incur comparable key-generation costs.

\begin{table}[!t]
  \caption{Circuit-specific key-generation time of VG-Halo2 and
  ZKGraph under the same Halo2/KZG backend (s).}
  \label{tab:keygen-comparison}
  \centering
  \footnotesize
  \setlength{\tabcolsep}{2pt}
  \setlength{\aboverulesep}{0.25ex}
  \setlength{\belowrulesep}{0.35ex}
  \renewcommand{\arraystretch}{1.02}
  \begin{tabular}{@{}ccccccccc@{}}
\toprule
Scale & System & IC1 & IC2 & IC9 & IC13 & IS3 & IS4 & IS5 \\
\midrule
\multirow{2}{*}{60k} & VG-Halo2 & 6.435 & 4.319 & 4.176 & 0.994 & 1.039 & 1.032 & 1.371 \\
 & ZKGraph & 6.225 & 5.322 & 5.556 & 1.424 & 2.141 & 1.604 & 1.822 \\
\midrule
\multirow{2}{*}{120k} & VG-Halo2 & 12.542 & 8.369 & 8.373 & 1.901 & 2.123 & 2.065 & 2.748 \\
 & ZKGraph & 11.724 & 10.603 & 11.176 & 2.716 & 4.276 & 3.176 & 3.660 \\
\midrule
\multirow{2}{*}{240k} & VG-Halo2 & 25.625 & 17.759 & 17.943 & 3.795 & 4.357 & 4.179 & 5.690 \\
 & ZKGraph & 23.798 & 22.330 & 23.273 & 5.622 & 9.111 & 6.651 & 7.707 \\
\bottomrule
\end{tabular}

\end{table}

\subsubsection*{Operator-Level Comparison}
\begin{figure}[!t]
  \centering
  \includegraphics[width=0.88\linewidth]{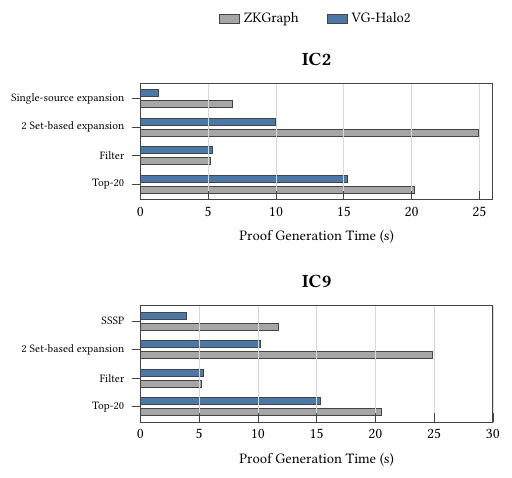}
  \caption{Operator-level proof-generation time of ZKGraph and
  VG-Halo2 for IC2 and IC9 at the 60k database scale.}
  \label{fig:operation-proof-time}
\end{figure}

We also decompose the IC2 and IC9 query circuits of both systems into
operator circuits at the 60k database scale. These two queries contain all
operators for which Section~\ref{sec:circuit-design} improves on ZKGraph's
proof constructions: single-source expansion and SSSP over bidirectional
edge relations, and set-based expansion. Because ZKGraph's public
implementation does not provide standalone circuits for all operators in IC2
and IC9, we construct each operator circuit from the code that implements the
operator in the complete query circuit, and we apply the same procedure to
VG-Halo2. We then
measure the proof-generation time of each operator. The experimental results
are shown in Figure~\ref{fig:operation-proof-time}.

VG-Halo2 reduces the proof-generation time of single-source expansion, SSSP,
and set-based expansion in both queries. Together with the system-level
comparison, these results show that our operator-proof
constructions improve performance even under ZKGraph's proving
backend.\footnote{VG-Halo2 uses the same filtering proof construction as
ZKGraph and exhibits similar proof-generation time for this operator.
VG-Halo2 also includes implementation-level constraint optimizations for
top-$k$, which contribute to its reduced proof-generation time.}

\section{Related Work}
\label{sec:related-work}

Verifiable query processing allows a client to check the correctness and
completeness of a result from an untrusted database service without
re-executing the query. Existing approaches can be grouped into
authenticated data structures, trusted hardware, and cryptographic proof
systems.

\subsubsection*{ADS-based methods}
Authenticated data structures (ADSs) bind a database or index to a compact
digest and return verification objects containing the index fragments needed to
check query answers. Dynamic authenticated indexes support updates and
freshness in outsourced databases~\cite{li2006dynamic}, while vChain develops
authenticated indexes for Boolean range queries over blockchain
data~\cite{xu2019vchain}. For graph data, authenticated hints support
shortest-path verification~\cite{yiu2010shortestpath}, and Merkle-based feature
indexes authenticate subgraph queries~\cite{fan2014authenticatedsubgraph}. TAP
further combines a tree-based ADS with zero-knowledge range proofs to verify
statistical queries~\cite{reijsbergen2023tap}. Although efficient for
index-supported lookups, ADS-based methods tie verification objects to specific
access paths and generally do not support confidential multi-hop graph queries.

\subsubsection*{Hardware-based methods}
Trusted hardware executes database operations within an attested enclave.
CorrectDB uses a tamper-resistant coprocessor to authenticate SQL
queries~\cite{bajaj2013correctdb}, and VeriDB builds verifiable relational
storage and SQL execution around SGX~\cite{zhou2021veridb}. These systems offer
low verification overhead and support expressive queries, but require trust in
hardware enclaves and need oblivious mechanisms to conceal data-dependent
access patterns.

\subsubsection*{Cryptographic proof methods}
IntegriDB and vSQL use cryptographic verifiable computation for SQL queries
over outsourced databases~\cite{zhang2015integridb,zhang2017vsql}, focusing on
result integrity rather than zero-knowledge privacy. To protect data privacy,
ZQL compiles simple queries over private data into zero-knowledge
protocols~\cite{fournet2013zql}, ZKSQL decomposes ad-hoc SQL queries into
operator-level interactive zero-knowledge protocols~\cite{li2023zksql}, and
PoneglyphDB uses optimized PLONKish circuits and recursive composition for
non-interactive SQL verification~\cite{gu2024poneglyphdb}. Recent systems
further co-design proofs with database execution: qedb combines algebraic table
commitments with modular query protocols for integrity
verification~\cite{botta2025qedb}, TruthTable composes polynomial protocols
with proof-aware query rewrites in DataFusion~\cite{namboothiry2026truthtable},
and V3DB proves a fixed-shape IVF-PQ pipeline for verifiable vector
search~\cite{qiu2026v3db}. However, these systems target relational or
vector-search pipelines rather than graph queries.

For graph processing, ALITHEIA develops tailored verification for graph
algorithms~\cite{zhang2014alitheia}. For decentralized graph data, VeriDKG uses
a blockchain-maintained ADS and a cryptographic accumulator to verify SPARQL
queries over knowledge graphs, whereas VGQ integrates an external graph
database to verify queries over blockchain transaction
graphs~\cite{zhou2023veridkg,yao2025vgq}. These systems focus on query
integrity rather than zero-knowledge evaluation over private graph data. Most
closely related to our work, ZKGraph introduces an expansion-centric
decomposition and corresponding PLONKish circuits as the first system for zero-knowledge graph
query evaluation~\cite{wu2025zkgraph}. Building on this model, VerfGraph
proposes new proof constructions for expansion and shortest-path operations to
reduce costly range checks, and introduces dedicated proofs for count and
weighted shortest-path operators, improving both proving efficiency and
query expressiveness.

\section{Conclusion}
\label{sec:conclusion}

We present VerfGraph, a zero-knowledge system for verifiable graph query
evaluation over committed graph data. Building on the expansion-centric model,
VerfGraph proposes new proof constructions for expansion and shortest-path
operations to reduce costly range checks and introduces dedicated proof
constructions for count and weighted shortest-path operators. Experiments on the LDBC SNB benchmark show
that VerfGraph substantially reduces proof-generation time and peak prover
memory usage compared with ZKGraph while maintaining millisecond-level
verification and sub-MB proofs.

\section*{AI-Generated Content Acknowledgement}

Generative AI tools were used solely to improve the clarity, conciseness,
grammar, and readability of the authors' original text. They were not used
to generate research ideas, design the methodology, conduct experiments,
analyze data, or draw conclusions. All AI-assisted edits were reviewed and
revised by the authors, who take full responsibility for the content of this
work.

\bibliographystyle{IEEEtran}
\bibliography{references}

\end{document}